\documentclass[a4paper]{jpconf}
\usepackage{graphicx}
\usepackage{amsmath}
\usepackage{subfigure}
\usepackage{times}
\usepackage{mathptmx}

\usepackage{fancyhdr}
\usepackage{ifpdf}
\ifpdf
\usepackage[pdftex]{hyperref}
\else
\usepackage[hypertex]{hyperref}
\fi

 \hypersetup{
   pdftitle={},%
   pdfauthor={J.P. Lansberg},%
   pdfsubject={},%
   pdfkeywords={},%
   pdfstartview={},%
   bookmarksopen=true, breaklinks=true, debug=true, %
   colorlinks=true, linkcolor=red, citecolor=blue, urlcolor=blue
 }

\newcommand{\vect}[1]{\vec{#1}}

\newcommand{\ie}{{\it i.e.}}
\newcommand{\eg}{{\rm e.g.}}
\newcommand{\eqs}[1]{\begin{equation} \begin{split} #1\end{split} \end{equation} }

\newcommand{\F}{\mathcal F}

\newcommand{\Q}{{\cal Q}}

\newcommand{\widebar}[1]{\overline{#1}}
\newcommand{\pp}{pp}

\def\sqrtsNN {\mbox{$\sqrt{s_{NN}}$}}

\def\RpPb    {\mbox{$R_{p\rm Pb}$}}
\def\RFB    {\mbox{$R_{\rm FB}$}}

\def\beq     {\begin{equation}}
\def\eeq     {\end{equation}}

\begin{document}
\title{Theory status of quarkonium production in
proton-nucleus collisions}

\author{J.P. Lansberg}

\address{Institut de Physique Nucl\'eaire, CNRS-IN2P3, Univ. Paris-Sud, Universit\'e Paris-Saclay, 91406~Orsay Cedex, France}

\ead{Jean-Philippe.Lansberg@in2p3.fr}

\begin{abstract}
I give a brief overview of the recent theoretical progress in the study of quarkonium production in proton-nucleus collisions in view of the recent LHC and RHIC results. A special emphasis is put on the excited states such as the $\psi'$, $\Upsilon(2S)$ and $\Upsilon(3S)$.
\end{abstract}

\section{On the importance of understanding conventional nuclear effects: \\ the CMS $\Upsilon(nS)$ sequential suppression example.}
\vskip 5pt

One of the highlights of the Run1-LHC heavy-ion results 
is admittedly the observation made by the CMS collaboration of a relative suppression of the $n=2,3$ $\Upsilon$ states with respect to the $n=1$ $\Upsilon$ state in PbPb collisions at $\sqrtsNN=2.76$ TeV. Such an observation was expected and even awaited for by the supporters of the phenomenon of ``sequential quarkonium suppression'' according to which 
the excited states with larger radii should suffer more from the colour screening in the deconfined matter~\cite{Matsui:1986dk}.

Actually, CMS published a set of two papers~\cite{Chatrchyan:2011pe,Chatrchyan:2012lxa} describing this observation, the second reinforcing the first by quoting measurements of individual nuclear modification factor\footnote{For minimum bias $AB$ collisions, 
such factors are defined such that $R_{AB}=\sigma_{AB}/(A B \sigma_{pp})$; they equal unity in absence of nuclear effects, that is when the nucleus-nucleus collision is like an incoherent superposition of nucleon-nucleon collisions.}. In discussing their results, they had in mind that other nuclear-matter effects than the creation of a quark-gluon plasma (QGP) could
suppress the $\Upsilon(nS)$, yet noting :
\begin{quote}\it
[...] such effects should have a small impact on the double ratios reported
here. Initial-state nuclear effects are expected to affect similarly each of the three $\Upsilon$
states, thereby canceling out in the ratio. Final-state “nuclear absorption” becomes
weaker with increasing energy and is expected to be negligible at the LHC.
\end{quote}
At this point, it is fair to say that nobody claimed the contrary and that this assertion was supported by a number of theoretical studies, \eg\ \cite{Vogt:2010aa,Ferreiro:2011xy,Arleo:2012rs}, where the excited states were expected to be suppressed as much as the $n=1$ state. In particular, such expectations were in line with the idea that any phenomenon taking place into the very fast moving lead nucleus (the nuclear matter in question) and which would
impact differently the  $\Upsilon$ states would {\it de facto} violate the Heisenberg principle. Indeed, the time they take to escape this nuclear
matter is way smaller, given the huge boost between their rest frame and the nucleus rest frame, than the time they need to form and to be distinguished.
The situation is obviously different if one considers effects resulting from a nucleus-nucleus collision, such as that of a QGP which is produced at small velocities/rapidities in the center-of-mass frame of the nucleus-nucleus reaction. This is precisely the region
 covered the LHC detectors which are used to detect quarkonia.

However, in 2013, the CMS collaboration carried out the corresponding analysis in proton-nucleus collisions
at 5 TeV~\cite{Chatrchyan:2013nza}. At the great surprise of many, they uncovered a significant relative 
suppression of the $2S$ and $3S$ states with respect to the $1S$ state. Quantitatively speaking, if 
the effects responsible for the relative $nS/1S$ suppression in $p$Pb collisions quoted in Table~\ref{tab:CMS-Upsilon} factorise (and is thus squared in the PbPb case), 
they could be responsible for half of the PbPb relative suppression strongly contradicting all previous expectations ! This
 illustrates how important state-of-the-art and dedicated studies in $p$Pb collisions are.

\begin{table}[t!]
\centering
\caption{Double ratios expressing the relative nuclear suppression of the excited $\Upsilon(2S,3S)$ states with respect 
to the $\Upsilon(1S)$ state as observed by CMS~\cite{Chatrchyan:2012lxa,Chatrchyan:2013nza} in PbPb collisions at $\sqrtsNN=2.76$ TeV and
in $p$Pb collisions at $\sqrtsNN=5$ TeV.}
\begin{tabular}{@{}*{7}{l}}
\br
$\frac{[\Upsilon\text{(nS)}/\Upsilon\text{(1S)}]_{ij}}{[\Upsilon\text{(nS)}/\Upsilon\text{(1S)}]_{\pp}}$ & $2S$ & $3S$\\
\mr 
PbPb& $0.21 \pm 0.07\,(\text{stat.}) \pm 0.02\,(\text{syst.})$ & $0.06 \pm 0.06\,(\text{stat.}) \pm 0.06\,(\text{syst.})$\\
$p$Pb& $0.83 \pm 0.05 \,(\text{stat.}) \pm 0.05\,(\text{syst.})$ & $0.71 \pm 0.08 \,(\text{stat.})\pm 0.09\,(\text{syst.})$\\
\br
\end{tabular}\label{tab:CMS-Upsilon}
\vspace*{-.5cm}
\end{table}

In addition to be of crucial importance to correctly interpret nucleus-nucleus observations -- as just illustrated, 
such proton-nucleus reactions involve many physics effects which are of high interest on their own and which can certainly 
tell us much about QCD at the interface between its pertubative and nonpertubative domain. They indeed provide means to study :
\begin{itemize}
\item[$\cdot$] the nuclear modification of the parton distributions of nucleons inside nuclei, also referred as to shadowing, antishadowing, EMC and Fermi-motion effects;
\item[$\cdot$] the phenomenon of gluon saturation and the high-energy limit of QCD, the so-called low-$x$ physics;
\item[$\cdot$] the time-evolution of a $Q\widebar Q$ pair and the dynamics of its hadronisation;
\item[$\cdot$] the propagation of quarks and gluons in a dense medium, the process of energy loss and the Cronin effect;
\item[$\cdot$] the quarkonium-production mechanisms, in particular the  colour octet vs. singlet contributions;    
\item[$\cdot$] the nonperturbative charm content of the proton;
\item[$\cdot$] the collinear-factorisation framework in the nuclear medium;
\item[$\cdot$] the quarkonium-hadron interactions;
\item[$\cdot$] the mechanisms underlying single-spin asymmetries, \dots 
\end{itemize}

In this proceedings contribution, I briefly review the current knowledge of quarkonium production in proton-nucleus
collisions at ultra-relativistic energies, with a specific focus on quarkonium radially-excited states.

\section{A baseline to understand the basics}
\vskip 5pt

Let us start by discussing a simple picture where only two effects are accounted for, 
namely (i) the inelastic scattering of the quarkonia with some nucleons in the nuclear matter, sometimes also called the nuclear
absorption
and (ii) the modification of the partonic densities in nucleons embedded in nuclei, historically
referred to as the EMC effect.

 The differential cross section as a function, for instance, of $y$, $P_T$ and $\vect b$ 
for the production of a quarkonium in a proton-nucleus collision
can be assumed to be obtained from the partonic one via the following convolution:  
\eqs{\frac{d \sigma_{pA\to \Q X}}{dy \, dP_T \, d\vect b }=
\!\int \!dx_1 dx_2  \, &  {g({x_1},\mu_F)} \int dz_A  
{\F^A_g({x_2},{\vect b}, z_B,\mu_F)} 
\\ & \times  2 \hat s P_T
\frac{d \sigma_{gg\to \Q+ g}}{d \hat t} 
\delta(\hat s-\hat t-\hat u-M^2)
{S_A({\vect b},z_A)}} 

The partonic differential cross section $\frac{d \sigma_{gg\to \Q+ g}}{d \hat t}$ 
can in principle be evaluated from any model (Colour Singlet, Colour Octet or Colour Evaporation Model for instance. See \cite{Andronic:2015wma,Brambilla:2010cs,Lansberg:2006dh} for reviews.) provided
that it  satisfactorily describes  the spectra under scrutiny, \ie\ $y$ or  $P_T$. 
To do so, we have developped a probabilistic Glauber Monte Carlo code, {\sc Jin}, which can handle any expression for $\frac{d \sigma_{gg\to \Q+ g}}{d \hat t}$~\cite{Ferreiro:2008qj,Ferreiro:2008wc,Ferreiro:2009ur}, yet accounting for the impact-parameter $\vect b$ dependence of any nuclear effect.

Since theoretical predictions
are usually compared to experimental measurements in the form of $R_{pA}$, the absolute normalisation of $\frac{d \sigma_{gg\to \Q+ g}}{d \hat t}$ cancels out. In general, its normalisation uncertainty (from the heavy-quark mass, the scales and  nonpertubative inputs), which can easily be on the order of a factor of 2 or 3, is significantly larger  than the expected nuclear effects, up to 50\% at best. It thus makes sense to keep looking at $R_{pA}$, rather than at $\sigma_{pA}$ not to be misguided by normalisation issues unrelated to nuclear effects.

\subsection{The so-called nuclear break-up}
\label{subsec:sigma_abs}

The first basic  effect and ingredient entering this simple approach 
is the  survival probability for a $Q\widebar{Q}$
produced at the point $(\vect r_A,z_A)$ to escape  
the nuclear medium unscathed. It is usually parametrised as
\eqs{S_A(\vect r_A, \,z_A)= \exp \left ( - A \,{\sigma_{\mathrm{break-up}}}
\int_{z_A}^{\infty} d\tilde{z}\ \rho_A(\vect r_A, \tilde{z}) \right ),}
by introducing $\sigma_{\mathrm{break-up}}$ as the cross section for 
inelastic collisions between a nucleon in the nuclear matter --the nucleus--
and the quarkonium ($\rho_A(\vect r_A, {z})$ is the nuclear density). Although it is sometimes meant to account for any effect
beyond those of the nuclear PDF (see below), $\sigma_{\mathrm{break-up}}$ is
nevertheless considered to be somehow connected to the size of the propagating object.
In particular,  {\it if} the meson is formed when it traverses the nuclear matter,
one expects $\sigma_{\mathrm{break-up}} \propto r^2_{\rm meson}$. In particular, 
$2S$ (and $3S$) states should significantly be more suppressed owing to their significantly larger radius.

In principle, $\sigma_{\mathrm{break-up}}$ is also connected to the meson-photoproduction cross section 
on nuclear target. This connection is however not always trivial~\cite{Benhar:1992df}.
 Both charmonia and bottomonia typically need $0.3 \div 0.4$ fm/$c$ to form -- in other words for one to be
 able to distinguish the $1S-2S$ energy levels in the meson rest frame. If the quarkonium
momentum in the nucleus rest frame increases, this time gets boosted and the states form way outside 
the nuclear matter. In such a case,  they are less suppressed and such a break-up mechanism should act 
on the same way on any state of a family. What propagates in the nuclear
matter is a sort of pre-resonant state whose quantum numbers are not yet determined.

To fix the idea, in the case of a quarkonium produced at $y=0$ in a $d$Au collision at RHIC at $\sqrt{s_{NN}}=200$ GeV, the boost between
the quarkonium rest frame and that of the gold nucleus\footnote{$\gamma(y=0) =\cosh(y_{\rm beam}=5.36)$.} is $\gamma = E_{\rm beam, cms}/m_N  \simeq 107$. 
It thus takes 30 fm/$c$ for a quarkonium to form and to become distinguishable from its excited states.
At the LHC (5 TeV), still for a particle with $y=0$, 
 $\gamma \simeq 2660$ ($y_{\rm beam}=8.58$).  The quarkonium-formation time seen from the nuclear-matter viewpoint is now
 $800 \div 1000$ fm/$c$. No nuclear-matter effect taking place over the size of the nucleus, \ie\ less than 15 fm, can in principle
act differently on a $J/\psi$ or $\Upsilon$ and its excited states. The boost at $y=0$ increases when one increases the collision energies and
obviously does so when one increases $y$ in the colliding nucleon direction. This formation-time effect has for instance
been invoked to explain the disappearance of the different $\psi'$ vs $J/\psi$ suppression measured by E866~\cite{Leitch:1999ea} at Fermilab when moving from $y\simeq 0$ ($x_F\simeq 0$) to forward rapidities ($x_F \to 1 $).

For extremely large values of $\gamma$, what propagates in the nucleus is probably a mere $Q\bar Q$ pair --whose colour state is however not known-- and, in such a case, one could
assume a naive high energy limit, $\sigma_{\mathrm{break-up}} \propto \pi/m_Q^2$, that is on the order of 0.5 mb for the charmonia. The situation is however much more complex, not only because of the suppression of the gluon densities, which I discuss below, but also because of non-linear 
coherent effects~\cite{Kopeliovich:2001ee} arising at high energies. 

In general, one admits that the ``conventional'' nuclear break-up is not among the dominant effects at LHC energies --except perhaps in the backward edge of the LHCb acceptance-- and can even be neglected. According to the above discussion about the formation time, what should matter to determine if $\sigma_{\mathrm{break-up}}$ is large or not
is $\sqrt{s_{\psi N}}$, or the energy of the quarkonium in the nucleus rest frame, rather than the collision energy, $\sqrt{s_{NN}}$. A global survey of data at $x_F \simeq 0$ however showed~\cite{Lourenco:2008sk} that $\sigma_{\mathrm{break-up}}$ seemed
to scale and to decrease with $\sqrt{s_{NN}}$. However, I stress that a scaling and a decrease with $\sqrt{s_{\psi N}}$ was not ruled out in presence of a strong gluon 
shadowing (see Fig. 11 (d) of~\cite{Lourenco:2008sk}) as I discuss now.

\subsection{Implementing nuclear PDFs and their impact-parameter dependences}

In addition to final-state interactions --once the meson is produced-- in the nuclear matter, 
one should indeed account for the nuclear 
modification of the initial parton densities. 
To do so, the nuclear gluon PDF and its spatial dependence, $\F^A_g(x_1,\vect r_A, z_A,\mu_F)$, 
are assumed to be factorisable~\cite{Klein:2003dj} in terms of the nucleon gluon PDFs : 
\eqs{\F^A_g(x_1,\vect r_A, z_A;\mu_F) = \rho_A(\vect r_A,z_A)   \times {g(x_1;\mu_F)} 
\times \left(1+\big[{R^A_g (x,\mu_F)}-1\big]N_{\rho_A}
\frac{\int dz \,\rho_A(\vect r_A,z)}{\int dz \,\rho_A(0,z)}\right)
\,}
where\footnote{$N_{\rho_A}$ is fixed such that 
$A^{-1} \int d^2\vect r_A \int dz_A \rho_A(\vect r_A,z_A) \left(1+\big[{R^A_g (x,\mu_F)}-1\big]N_{\rho_A}
\frac{\int dz \,\rho_A(\vect r_A,z)}{\int dz \,\rho_A(0,z)}\right) = R^A_g (x,\mu_F)$.} $R^A_g (x,\mu_F)$ is the ratio of the gluon density per nucleon in a nucleus $A$ by
that in a free proton at a momentum fraction $x$ and a factorisation scale $\mu_F$. When discussing $R^A_g (x,\mu_F)$, 4 regions are distinguished depending on the value of $x$, 
namely the (i) Fermi-motion region ($x>0.7$), (ii) the EMC region ($0.3< x<0.7$), (iii) the anti-shadowing region ($0.05<x<0.3$), (iv) the shadowing region ($x < 0.05$). As what regards the gluons, I stress that only the shadowing depletion is established but its magnitude is still discussed. The gluon antishadowing has not yet been observed although assumed in many studies. For instance, 
it is absent in some nPDF fits~\cite{deFlorian:2003qf}. The gluon EMC effect is even less known. We have claimed in~\cite{Ferreiro:2011xy}
that the backward $\Upsilon$ data at RHIC may hint at a significant gluon EMC effect, perhaps stronger
than for quarks.

\subsection{Overall}

\begin{figure}[t!]
\centering 
\caption{\label{Fig:RpA-LHC}\RpPb\ at $\sqrtsNN=5$ TeV: typical comparisons between the ALICE and LHCb data and the effect induced by the gluon shadowing only (see text).}
\subfigure[$J/\psi$]{\includegraphics[width=0.45\columnwidth]{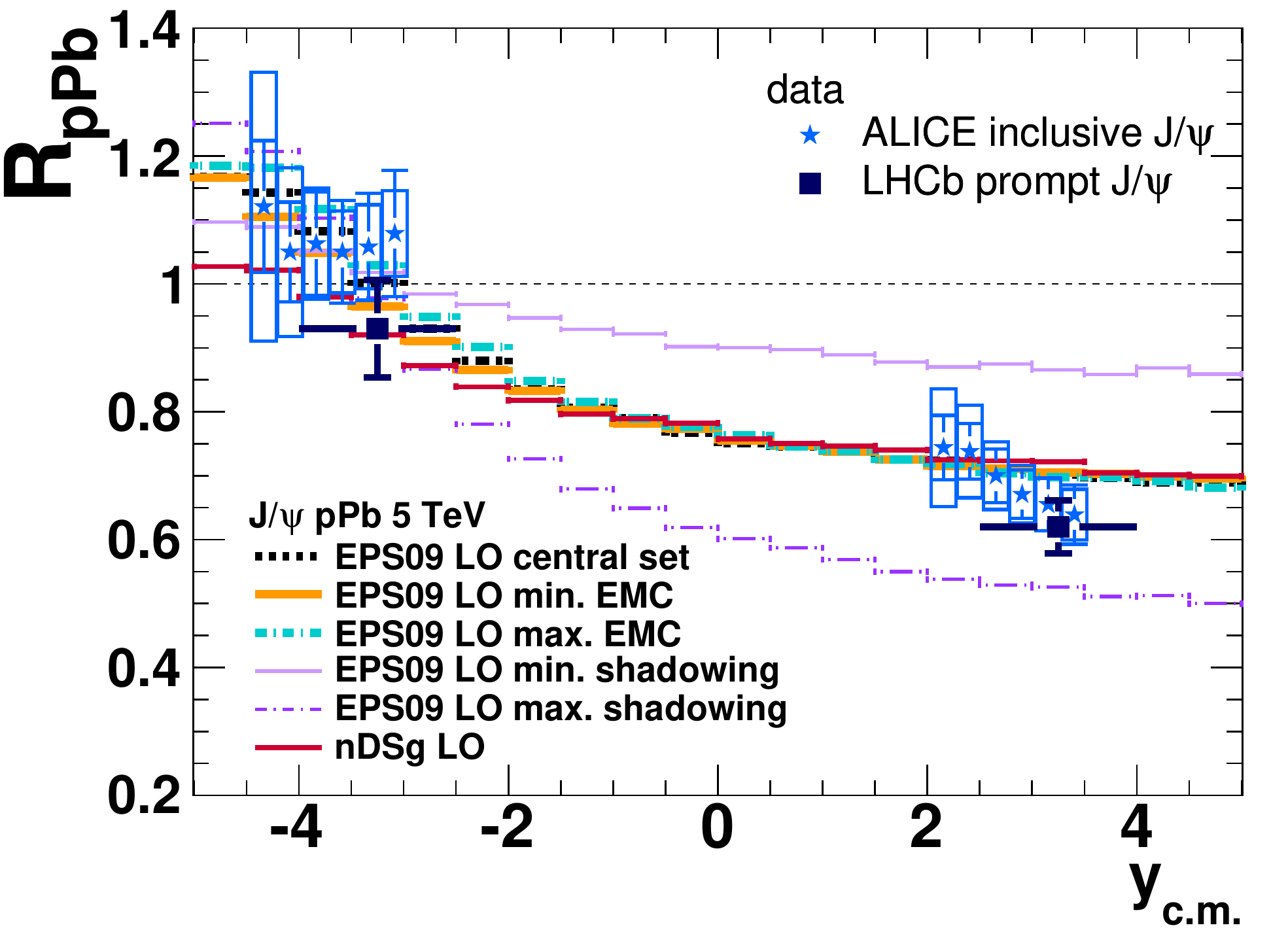}}
\subfigure[$\Upsilon(1S)$]{\includegraphics[width=.45\columnwidth]{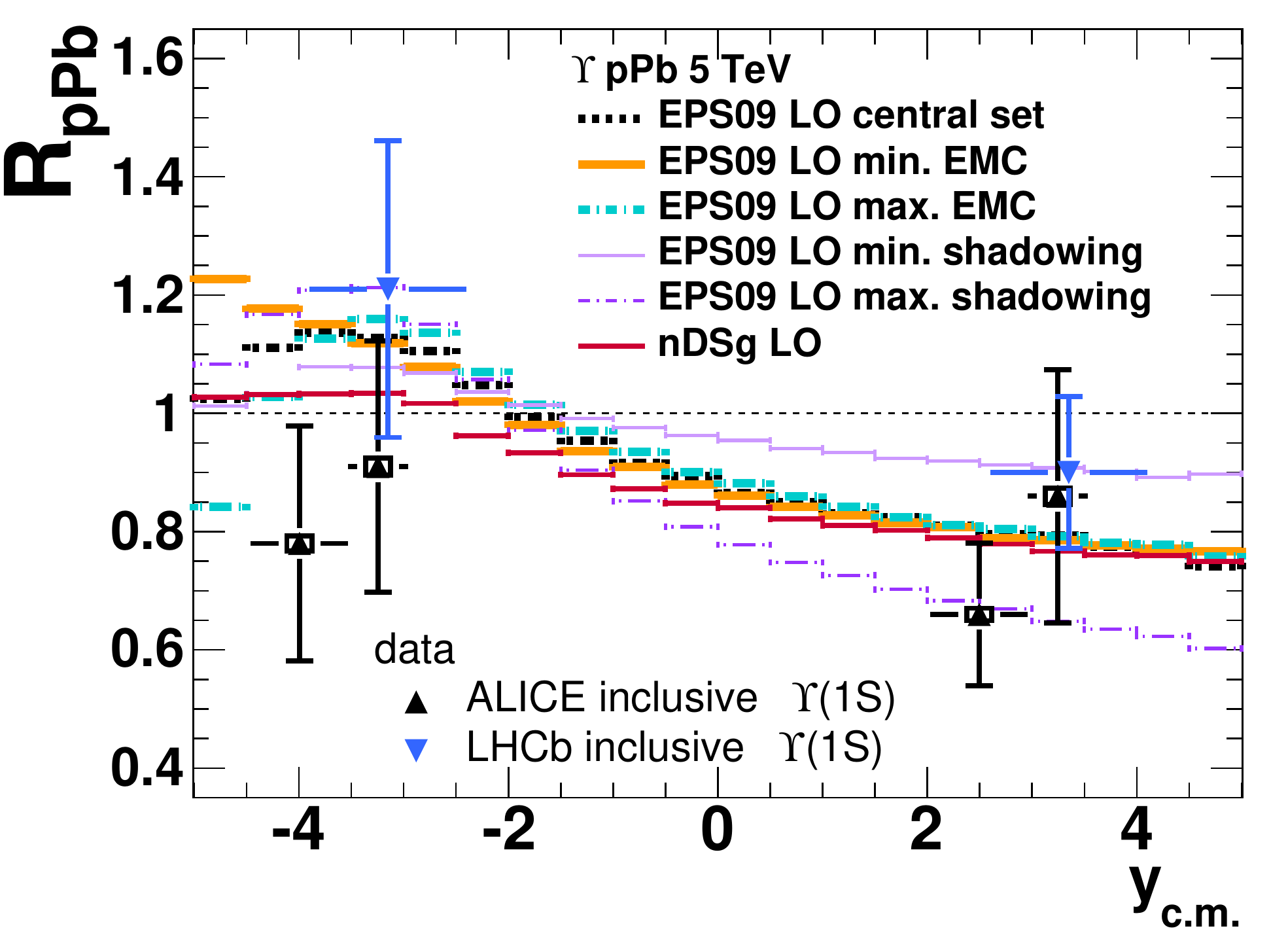}}
\vspace*{-.75cm}
\end{figure}

Typical comparisons at LHC energies when the break-up is neglected are shown on Figure~\ref{Fig:RpA-LHC}. A few comments are in order: (i) as discussed in~\cite{Ferreiro:2013pua,Ferreiro:2011xy}, such evaluations do depend on the factorisation scale as well as on the nPDF set chosen\footnote{In addition to EPS09~\cite{Eskola:2009uj} and nDSg~\cite{deFlorian:2003qf} used here, other common nPDFs sets which could have been used are nCTEQ15\cite{Kovarik:2015cma}, HKN~\cite{Hirai:2007sx}, DSSZ\cite{deFlorian:2011fp}, FSG\cite{Frankfurt:2011cs}, ...}; The corresponding uncertainties are not shown and can be as large as than of a given nPDF set; (ii) the global normalisation uncertainties of the data due to the absence of $\pp$ data at 5 TeV are also not shown. These however cancel in $\RFB$ (see~\cite{Ferreiro:2013pua}). (iii) such a computation accounts for a $2\to 2$ parton scattering. Evaluations with a simplified $2\to 1$ kinematics and based on the colour evaporation model give similar results~\cite{Vogt:2010aa,Vogt:2015uba} (See~\cite{Ferreiro:2009ur} for a detailed discussion).

RHIC results have been discussed at length within this simple picture in \eg~\cite{Ferreiro:2008wc,Ferreiro:2009ur,Ferreiro:2012sy,McGlinchey:2012bp}. $\sigma_{\mathrm{break-up}}$ was fitted to the PHENIX data. In general, this method works except in some parts of the phase space: 
a counter-intuitive increase of $\sigma_{\mathrm{break-up}}$ at forward rapidities\footnote{except in presence of
a strong gluon shadowing~\cite{Ferreiro:2009ur}.}, a slight $p_T$ enhancement difficult 
to explain at backward rapidites, a hint at a different centrality dependence as compared to nPDFs proportional to the local
nuclear density, \dots\ For more details, I refer the interested reader to the recent review~\cite{Andronic:2015wma} and references therein.
\section{Going further by adding some nuclear effects}
\vskip 5pt

Whereas the simple model discussed above probably accounts reasonably well for initial state effects via nuclear PDFs, the treatment of the final state effects is indeed rather empirical for $\sigma_{\mathrm{break-up}}$ 
is fitted to the data and its value {\it a posteriori} confronted
to qualitative expectations. As just said, it works but some ``anomalies'' have lately been observed.

This has motivated more advanced theoretical studies. In particular, it was
argued~\cite{Arleo:2010rb} that a coherent energy loss, scaling like the projectile energy, could explain well the data from fixed-target to LHC energies. In presence of such an energy loss, arising from interferences between initial and final state radiations, 
no additional break-up is needed. In a sense, it accounts for it. However, it is not clear if gluon shadowing is needed when
such an energy loss is ``switched on''.

The treatment of small-$x$ effects like the gluon saturation, which were initially predicted~\cite{Fujii:2013gxa} 
to be stronger than the expected gluon-shadowing suppression from global nPDF fits, have been improved. 
Current postdictions~\cite{Ducloue:2015gfa} are in better agreement with data\footnote{It has also 
recently been claimed that the suppression may essentially depend on the quantum state 
of propagating pair~\cite{Ma:2015sia}, emphasising that such a small-$x$ suppression is not merely an initial state effect.}. They are thus also 
on the same order as nPDF predictions and it is legitimate to wonder if different physics is really at work.

Finally, it was recently noted in~\cite{Andronic:2015wma} that the nuclear suppression seen in the data seems not to show an increase from RHIC to LHC energies as most of the models had predicted it : higher energy, smaller $x$, stronger shadowing. Yet, it seems that the nuclear suppression decreases when progressively going to heavier systems, \ie\ from $J/\psi$ to $\Upsilon$, passing by  $b(\to J/\psi)$.

 \section{The excited state puzzle(s)}
\vskip 5pt

As I have reported in the introduction, CMS uncovered an unexpected relative suppression of the $\Upsilon(2S,3S)$ state to
the $\Upsilon(1S)$. As it can be read on Table~\ref{tab:CMS-Upsilon}, it is on the order 20\% for the $\Upsilon(2S)$ 
and 30\% for the $\Upsilon(3S)$. Along the lines of the discussion of section \ref{subsec:sigma_abs}, it is extremely 
counter-intuitive because of the very large boost between the quarkonia and the nucleus at LHC energies.

\begin{figure}[t!]
\centering 
\caption{\label{Fig:psip-jpsi} Ratios expressing the relative nuclear suppression of the $\psi'$ over the $J/\psi$
at  $\sqrtsNN=200$ GeV in $d$Au collisions and at $\sqrtsNN=5$ TeV in $p$Pb collisions: comparisons between the PHENIX~\cite{Adare:2013ezl} and ALICE~\cite{Abelev:2014zpa} data and the results of the comover interaction model~\cite{Ferreiro:2014bia}. 
[A 28\% systematical normalisation uncertainty on the PHENIX data is not shown]. Plot from~\cite{Andronic:2015wma}.}
\includegraphics[width=0.8\columnwidth]{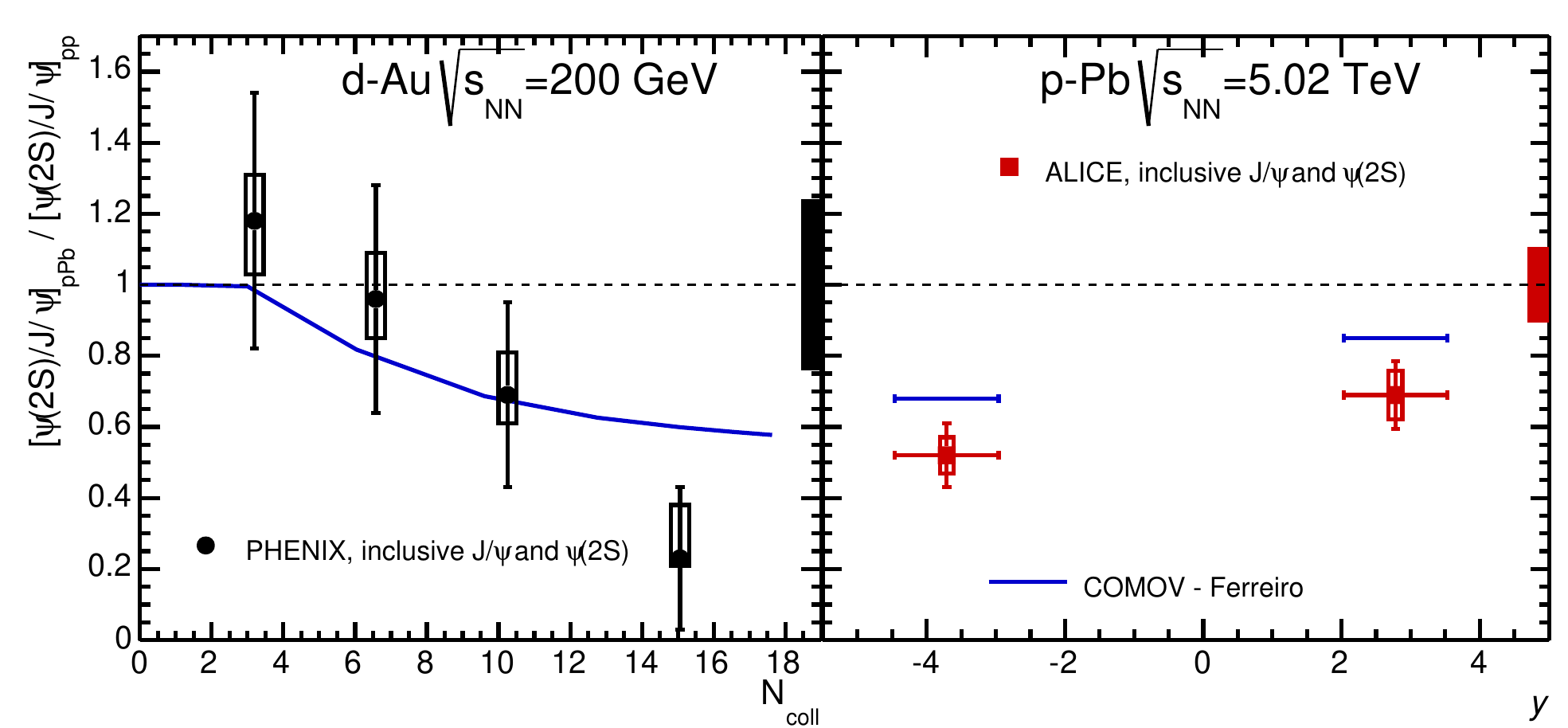}
\vspace*{-.5cm}
\end{figure}

In fact, a similar observation was also made by the PHENIX experiment at 200 GeV. They reported~\cite{Adare:2013ezl}
a relative suppression of the $\psi'$ to the $J/\psi$ increasing with the centrality 
(but with a large ($\simeq 25\%$) global systematical uncertainty). Integrating over centrality, the ratio
$[\psi'/J/\psi]_{d\rm Au}/[\psi'/J/\psi]_{\pp}=0.68 \pm 0.14 ^{+0.13}_{-0.08} \pm 0.16 $, \ie\ smaller than unity but with significant uncertainties. Following this analysis, we confirmed~\cite{Ferreiro:2012mm} that the theoretical uncertainties in the nPDF impact evaluation (mass in the kinematics, scale) is at most 2 \%, \ie\ much smaller.
ALICE also performed an analysis~\cite{Abelev:2014zpa} in $p$Pb collisions at 5 TeV (see Figure \ref{Fig:psip-jpsi} (right panel))
and also uncovered a relative suppression of larger magnitude in the backward region that in the forward region.  

As for now, the only approach which can describe these results without tuning parameters\footnote{In~\cite{Du:2015wha}, Du and Rapp 
studied the impact of some of their model parameters on the dAu RHIC relative suppression 
which is then used to study the PbPb LHC relative suppression.} is the comover interaction model (CIM) (see~\cite{Ferreiro:2012rq} and references therein). Such an approach accounts for final-state scatterings with comoving hadrons along
the quarkonium trajectory. These can occur over times long enough for the quarkonium to be formed; differences
between the $1S$ and $2S$ suppression can thus occur. As for the break-up cross section, one introduces an empirical comover-quarkonium
inelastic cross section which is however fixed to reproduce $AA$ data where the comover density is very large. 
Such a picture also naturally explains a larger relative suppression in the nucleus rapidity region (negative $y$ at the LHC) where
the comover density --connected to the charged particle multiplicity-- is larger. 
Comparisons between the CIM results and data from PHENIX and ALICE are shown on Figure \ref{Fig:psip-jpsi}.

 \section{Conclusions and outlooks}
\vskip 5pt

As I argued, the study of quarkonium production in proton-nucleus collisions is of fundamental importance if one wishes to 
properly interpret --quantitatively but even also qualitatively-- the physics underlying the quarkonium suppression in nucleus-nucleus collisions. The case of the sequential suppression of $\Upsilon$ measured by CMS is a prime example. Just measuring
quarkonium production in nucleus-nucleus collisions is not enough !

Moreover, such studies are of highest relevance for themselves. It is very important to identify the mechanisms responsible 
for the relative suppression of quarkonium excited states seen at RHIC and the LHC. Indeed, at high energies, final-state 
interactions within the nucleus --as the ones included in the conventional models discussed here-- occur too early to impact differently the different energy levels. One of the mechanisms which can naturally explain such a relative suppression is the possibility of  scatterings over ``long'' times with comoving hadrons. This is included in the comover interaction model which was introduced long ago to explain part of the anomalous quarkonium suppression in $AA$ collisions. 

As outlooks, let us stress that LHCb recently reported preliminary results on $\psi'$ production in $p$Pb collisions~\cite{LHCb-CONF-2015-005}. 
Their measurement confirms the ALICE one reported above. It also has the potential to extend it with the extraction 
of the (relative) suppression of non-prompt $\psi'$.  On general grounds, $\psi'$ from $b$ come from a process at a higher scale where some nuclear 
effect may be smaller but, more importantly, such $\psi'$ are produced a fraction of a millimeter --as opposed to femtometers--
after the collisions. As such, it is hard to believe that they would suffer from any nuclear final-state effect. 
As for now, the LHCb uncertainty is somewhat too large to draw any strong conclusion, 
the non-prompt $\psi'$ suppression is both compatible with that of non-prompt $J/\psi$ 
--which would be compatible with the absence of final-state interactions-- but also with that of prompt $\psi'$, which would be another stunning result.
As stunning as it could be, such a possibility should however not be ruled out given the 
previous surprising observations reported here which clearly challenged the established ideas.

In the longer run, let us add that ATLAS has also recently reported $J/\psi$ measurements in $p$Pb collisions~\cite{Aad:2015ddl} 
at finite $P_T$ and could also probably add some information on the excited state sector. Along these lines, 
the $P_T$ dependence of the relative $\psi'$ over $J/\psi$ suppression has still to be addressed theoretically and experimental uncertainties
can certainly be reduced.
Complementary measurements by LHCb in $p$Pb collisions bearing on Drell-Yan and open charm production may also shed some light
on some nuclear-matter effects at work in the same kinematical region as for quarkonia. The same also holds for measurements
in the fixed-target mode at the LHC, in the energy range of RHIC, but at negative $x_F$ and with unprecedented
luminosities~\cite{Massacrier:2015qba,Arleo:2015lja,vogtAFTER}. For an overview, 
I guide the reader to~\cite{Brodsky:2012vg,Lansberg:2012kf}.

To complete the picture, let us mention the importance of looking at quarkonium-associated production in $p$Pb collisions
which could uncover a new geometric scaling regime $A^{3/2}$ vs $A^1$ \cite{Strikman:2001gz,d'Enterria:2014bga}.
Recent theoretical~\cite{Artoisenet:2007xi,Lansberg:2009db,Gong:2012ah,Lansberg:2013wva,Lansberg:2013qka,Lansberg:2014swa,Lansberg:2015lva} and experimental~\cite{Aaij:2011yc,Aaij:2012dz,Aad:2014kba,Aad:2014rua,Khachatryan:2014iia} studies in $pp$ collisions paved the way for such studies.

\vspace*{-.5cm}
\ack 
I would like to thank A. Rakotazafindrabe for the plots of Figure 1 as well as the SQM organisers, in particular D. Blaschke and J. Aichelin for the kind invitation to present this overview. This work is supported in part by the CNRS (D\'efi Inphyniti -- Th\'eorie LHC France).

\vskip 5pt

\bibliographystyle{iopart-num}

\bibliography{proc_SQM2015}

\providecommand{\newblock}{}
\begin{thebibliography}{10}
\expandafter\ifx\csname url\endcsname\relax
  \def\url#1{{\tt #1}}\fi
\expandafter\ifx\csname urlprefix\endcsname\relax\def\urlprefix{URL }\fi
\providecommand{\eprint}[2][]{\url{#2}}

\bibitem{Matsui:1986dk}
Matsui T and Satz H 1986 {\em Phys. Lett.\/} {\bf B178} 416

\bibitem{Chatrchyan:2011pe}
Chatrchyan S {\em et~al.\/} (CMS) 2011 {\em Phys. Rev. Lett.\/} {\bf 107}
  052302 (\textit{Preprint} \eprint{1105.4894})

\bibitem{Chatrchyan:2012lxa}
Chatrchyan S {\em et~al.\/} (CMS) 2012 {\em Phys. Rev. Lett.\/} {\bf 109}
  222301 (\textit{Preprint} \eprint{1208.2826})

\bibitem{Vogt:2010aa}
Vogt R 2010 {\em Phys. Rev.\/} {\bf C81} 044903 (\textit{Preprint}
  \eprint{1003.3497})

\bibitem{Ferreiro:2011xy}
Ferreiro E~G, Fleuret F, Lansberg J~P, Matagne N and Rakotozafindrabe A 2013
  {\em Eur. Phys. J.\/} {\bf C73} 2427 (\textit{Preprint} \eprint{1110.5047})

\bibitem{Arleo:2012rs}
Arleo F and Peigne S 2013 {\em JHEP\/} {\bf 03} 122 (\textit{Preprint}
  \eprint{1212.0434})

\bibitem{Chatrchyan:2013nza}
Chatrchyan S {\em et~al.\/} (CMS) 2014 {\em JHEP\/} {\bf 04} 103
  (\textit{Preprint} \eprint{1312.6300})

\bibitem{Andronic:2015wma}
Andronic A {\em et~al.\/} 2015  (\textit{Preprint} \eprint{1506.03981})

\bibitem{Brambilla:2010cs}
Brambilla N {\em et~al.\/} 2011 {\em Eur. Phys. J.\/} {\bf C71} 1534
  (\textit{Preprint} \eprint{1010.5827})

\bibitem{Lansberg:2006dh}
Lansberg J~P 2006 {\em Int. J. Mod. Phys.\/} {\bf A21} 3857--3916
  (\textit{Preprint} \eprint{hep-ph/0602091})

\bibitem{Ferreiro:2008qj}
Ferreiro E~G, Fleuret F and Rakotozafindrabe A 2009 {\em Eur. Phys. J.\/} {\bf
  C61} 859--864 (\textit{Preprint} \eprint{0801.4949})

\bibitem{Ferreiro:2008wc}
Ferreiro E~G, Fleuret F, Lansberg J~P and Rakotozafindrabe A 2009 {\em Phys.
  Lett.\/} {\bf B680} 50--55 (\textit{Preprint} \eprint{0809.4684})

\bibitem{Ferreiro:2009ur}
Ferreiro E~G, Fleuret F, Lansberg J~P and Rakotozafindrabe A 2010 {\em Phys.
  Rev.\/} {\bf C81} 064911 (\textit{Preprint} \eprint{0912.4498})

\bibitem{Benhar:1992df}
Benhar O, Kopeliovich B~Z, Mariotti C, Nicolaev N~N and Zakharov B~G 1992 {\em
  Phys. Rev. Lett.\/} {\bf 69} 1156--1159

\bibitem{Leitch:1999ea}
Leitch M~J {\em et~al.\/} (NuSea) 2000 {\em Phys. Rev. Lett.\/} {\bf 84}
  3256--3260 (\textit{Preprint} \eprint{nucl-ex/9909007})

\bibitem{Kopeliovich:2001ee}
Kopeliovich B, Tarasov A and Hufner J 2001 {\em Nucl. Phys.\/} {\bf A696}
  669--714 (\textit{Preprint} \eprint{hep-ph/0104256})

\bibitem{Lourenco:2008sk}
Lourenco C, Vogt R and Woehri H~K 2009 {\em JHEP\/} {\bf 02} 014
  (\textit{Preprint} \eprint{0901.3054})

\bibitem{Klein:2003dj}
Klein S~R and Vogt R 2003 {\em Phys. Rev. Lett.\/} {\bf 91} 142301
  (\textit{Preprint} \eprint{nucl-th/0305046})

\bibitem{deFlorian:2003qf}
de~Florian D and Sassot R 2004 {\em Phys. Rev.\/} {\bf D69} 074028
  (\textit{Preprint} \eprint{hep-ph/0311227})

\bibitem{Ferreiro:2013pua}
Ferreiro E~G, Fleuret F, Lansberg J~P and Rakotozafindrabe A 2013 {\em Phys.
  Rev.\/} {\bf C88} 047901 (\textit{Preprint} \eprint{1305.4569})

\bibitem{Eskola:2009uj}
Eskola K~J, Paukkunen H and Salgado C~A 2009 {\em JHEP\/} {\bf 04} 065
  (\textit{Preprint} \eprint{0902.4154})

\bibitem{Kovarik:2015cma}
Kovarik K {\em et~al.\/} 2015  (\textit{Preprint} \eprint{1509.00792})

\bibitem{Hirai:2007sx}
Hirai M, Kumano S and Nagai T~H 2007 {\em Phys. Rev.\/} {\bf C76} 065207
  (\textit{Preprint} \eprint{0709.3038})

\bibitem{deFlorian:2011fp}
de~Florian D, Sassot R, Zurita P and Stratmann M 2012 {\em Phys. Rev.\/} {\bf
  D85} 074028 (\textit{Preprint} \eprint{1112.6324})

\bibitem{Frankfurt:2011cs}
Frankfurt L, Guzey V and Strikman M 2012 {\em Phys. Rept.\/} {\bf 512} 255--393
  (\textit{Preprint} \eprint{1106.2091})

\bibitem{Vogt:2015uba}
Vogt R 2015 {\em Phys. Rev.\/} {\bf C92} 034909 (\textit{Preprint}
  \eprint{1507.04418})

\bibitem{Ferreiro:2012sy}
Ferreiro E~G, Fleuret F, Lansberg J~P, Matagne N and Rakotozafindrabe A 2012
  {\em Few Body Syst.\/} {\bf 53} 27--36 (\textit{Preprint} \eprint{1201.5574})

\bibitem{McGlinchey:2012bp}
McGlinchey D~C, Frawley A~D and Vogt R 2013 {\em Phys. Rev.\/} {\bf C87} 054910
  (\textit{Preprint} \eprint{1208.2667})

\bibitem{Arleo:2010rb}
Arleo F, Peigne S and Sami T 2011 {\em Phys. Rev.\/} {\bf D83} 114036
  (\textit{Preprint} \eprint{1006.0818})

\bibitem{Fujii:2013gxa}
Fujii H and Watanabe K 2013 {\em Nucl. Phys.\/} {\bf A915} 1--23
  (\textit{Preprint} \eprint{1304.2221})

\bibitem{Ducloue:2015gfa}
Duclou\'e B, Lappi T and Mantysaari H 2015 {\em Phys. Rev.\/} {\bf D91} 114005
  (\textit{Preprint} \eprint{1503.02789})

\bibitem{Ma:2015sia}
Ma Y~Q, Venugopalan R and Zhang H~F 2015 {\em Phys. Rev.\/} {\bf D92} 071901
  (\textit{Preprint} \eprint{1503.07772})

\bibitem{Adare:2013ezl}
Adare A {\em et~al.\/} (PHENIX) 2013 {\em Phys. Rev. Lett.\/} {\bf 111} 202301
  (\textit{Preprint} \eprint{1305.5516})

\bibitem{Ferreiro:2012mm}
Ferreiro E~G, Fleuret F, Lansberg J~P and Rakotozafindrabe A 2012  [J. Phys.
  Conf. Ser.422,012018(2013)] (\textit{Preprint} \eprint{1211.4749})

\bibitem{Abelev:2014zpa}
Abelev B~B {\em et~al.\/} (ALICE) 2014 {\em JHEP\/} {\bf 12} 073
  (\textit{Preprint} \eprint{1405.3796})

\bibitem{Ferreiro:2014bia}
Ferreiro E~G 2015 {\em Phys. Lett.\/} {\bf B749} 98--103 (\textit{Preprint}
  \eprint{1411.0549})

\bibitem{Du:2015wha}
Du X and Rapp R 2015 {\em Nucl. Phys.\/} {\bf A943} 147--158 (\textit{Preprint}
  \eprint{1504.00670})

\bibitem{Ferreiro:2012rq}
Ferreiro E~G 2014 {\em Phys. Lett.\/} {\bf B731} 57--63 (\textit{Preprint}
  \eprint{1210.3209})

\bibitem{LHCb-CONF-2015-005}
Yang Z LHCb-CONF-2015-005 {\em ~\/} {\bf ~}

\bibitem{Aad:2015ddl}
Aad G {\em et~al.\/} (ATLAS) 2015 {\em Phys. Rev.\/} {\bf C92} 034904
  (\textit{Preprint} \eprint{1505.08141})

\bibitem{Massacrier:2015qba}
Massacrier L, Trzeciak B, Fleuret F, Hadjidakis C, Kikola D, Lansberg J~P and
  Shao H~S 2015 {\em Adv. High Energy Phys.\/} {\bf 2015} 986348
  (\textit{Preprint} \eprint{1504.05145})

\bibitem{Arleo:2015lja}
Arleo F and Peign\'e S 2015 {\em Adv. High Energy Phys.\/} {\bf 2015} 961951
  (\textit{Preprint} \eprint{1504.07428})

\bibitem{vogtAFTER}
Vogt R 2015 {\em Adv. High Energy Phys.\/} {\bf 2015} 492302

\bibitem{Brodsky:2012vg}
Brodsky S~J, Fleuret F, Hadjidakis C and Lansberg J~P 2013 {\em Phys. Rept.\/}
  {\bf 522} 239--255 (\textit{Preprint} \eprint{1202.6585})

\bibitem{Lansberg:2012kf}
Lansberg J~P, Brodsky S~J, Fleuret F and Hadjidakis C 2012 {\em Few Body
  Syst.\/} {\bf 53} 11--25 (\textit{Preprint} \eprint{1204.5793})

\bibitem{Strikman:2001gz}
Strikman M and Treleani D 2002 {\em Phys. Rev. Lett.\/} {\bf 88} 031801
  (\textit{Preprint} \eprint{hep-ph/0111468})

\bibitem{d'Enterria:2014bga}
d'Enterria D and Snigirev A~M 2014 {\em Nucl. Phys. A\/} {\bf 932} 296–301
  (\textit{Preprint} \eprint{1403.2335})

\bibitem{Artoisenet:2007xi}
Artoisenet P, Lansberg J~P and Maltoni F 2007 {\em Phys. Lett.\/} {\bf B653}
  60--66 (\textit{Preprint} \eprint{hep-ph/0703129})

\bibitem{Lansberg:2009db}
Lansberg J~P 2009 {\em Phys. Lett.\/} {\bf B679} 340--346 (\textit{Preprint}
  \eprint{0901.4777})

\bibitem{Gong:2012ah}
Gong B, Lansberg J~P, Lorce C and Wang J 2013 {\em JHEP\/} {\bf 03} 115
  (\textit{Preprint} \eprint{1210.2430})

\bibitem{Lansberg:2013wva}
Lansberg J~P and Lorce C 2013 {\em Phys. Lett.\/} {\bf B726} 218--222 [Erratum:
  Phys. Lett.B738,529(2014)] (\textit{Preprint} \eprint{1303.5327})

\bibitem{Lansberg:2013qka}
Lansberg J~P and Shao H~S 2013 {\em Phys. Rev. Lett.\/} {\bf 111} 122001
  (\textit{Preprint} \eprint{1308.0474})

\bibitem{Lansberg:2014swa}
Lansberg J~P and Shao H~S 2014  (\textit{Preprint} \eprint{1410.8822})

\bibitem{Lansberg:2015lva}
Lansberg J~P and Shao H~S 2015 {\em Nucl. Phys. B\/} (\textit{Preprint}
  \eprint{1504.06531})

\bibitem{Aaij:2011yc}
Aaij R {\em et~al.\/} (LHCb) 2012 {\em Phys. Lett.\/} {\bf B707} 52--59
  (\textit{Preprint} \eprint{1109.0963})

\bibitem{Aaij:2012dz}
Aaij R {\em et~al.\/} (LHCb) 2012 {\em JHEP\/} {\bf 06} 141 [Addendum:
  JHEP03,108(2014)] (\textit{Preprint} \eprint{1205.0975})

\bibitem{Aad:2014kba}
Aad G {\em et~al.\/} (ATLAS) 2015 {\em Eur. Phys. J.\/} {\bf C75} 229
  (\textit{Preprint} \eprint{1412.6428})

\bibitem{Aad:2014rua}
Aad G {\em et~al.\/} (ATLAS) 2014 {\em JHEP\/} {\bf 04} 172 (\textit{Preprint}
  \eprint{1401.2831})

\bibitem{Khachatryan:2014iia}
Khachatryan V {\em et~al.\/} (CMS) 2014 {\em JHEP\/} {\bf 09} 094
  (\textit{Preprint} \eprint{1406.0484})

\end{thebibliography}

\end{document}